# Material-Dependencies of the THz Emission from Plasmonic Graphene-Based Photoconductive Antenna Structures

Christoph Suessmeier[1], Sergi Abadal[3], Daniel Stock[1], Stephan Schaeffer[1], Eduard Alarcón[3],
Seyed Ehsan Hosseininejad[3], Anna Katharina Wigger[1], Stefan Wagner[2], Albert Cabellos-Aparicio[3],
Max Lemme[2] and Peter Haring Bolívar[1]

[1]Institute for High Frequency and Quantum Electronics, University of Siegen, 57076 Siegen, Germany
[2]Faculty of Electrical Engineering and Information Technology, RWTH Aachen University, 52074 Aachen, Germany
[3]NaNoNetworking Center in Catalunya (N3Cat), Universitat Politècnica de Catalunya, 08034 Barcelona, Spain

*Abstract*— Graphene supports surface plasmon polaritons with comparatively slow propagation velocities in the THz region, which becomes increasingly interesting for future communication technologies. This ability can be used to realize compact antennas, which are up to two orders of magnitude smaller than their metallic counterparts. For a proper functionality of these antennas some minimum material requirements have to be fulfilled, which are presently difficult to achieve, since the fabrication and transfer technologies for graphene are still evolving. In this work we analyze available graphene materials experimentally and extract intrinsic characteristics at THz frequencies, in order to predict the dependency of the THz signal emission threshold as a function of the graphene relaxation time $\tau_r$ and the chemical potential $\mu_c$.

## I. INTRODUCTION

THE decreasing size of miniaturized components such as physical, biological and chemical sensors leads to an increasing demand for compact and on-chip communication techniques, like wireless data transfer via miniaturized antennas [1]. Scaling classical metallic antennas to the size of several micrometers imposes the use of high frequencies in the near infrared and visible spectrum, where the conductivity has a complex frequency dependence [2]. Graphene is predestined as an alternative, because of the dispersion relation and its ability to support surface-plasmon polaritons (SPPs). The special dispersion relation induced plasmonic resonance of graphene is in the THz range [3], and thus of increasing interest in short-range communications [4]. Graphene excels its metallic counterparts since the antennas are up to two orders of magnitude smaller in area, as predicted by simulations [5]. Material parameters such as the relaxation time $\tau_r$ and the chemical potential $\mu_c$, are crucial for the adequate functionality (e.g. efficiency, gain, impedance matching, …) of such antennas

For an experimentally validated assessment of this potential, a closer analysis of the minimum required graphene material parameters is required. Experimental material parameters and dipolar emission characteristics are evaluated, simulating material influences on the threshold of the THz emission.

## II. RESULTS

The THz conductivity of the material is a basic parameter for the functionality of an antenna and can be described with a Drude model following the random phase approximation [6].

$$\sigma(\omega_r) = 2\frac{e^2}{\pi\hbar}\frac{k_B T}{\hbar}\ln\left[2\cosh\left[\frac{\mu_C}{2\,k_B T}\right]\right]\frac{i}{\omega_r + i\,\tau_r^{-1}} \quad (1)$$

with the elementary charge $e$, the reduced Planck constant $\hbar$, the Boltzmann constant $k_B$ and the temperature $T$.

Thus the conductivity of graphene strongly depends on the chemical potential and the relaxation time. In our approach a direct experimental assessment of these quantities is adopted, together with a simulation of the minimum required graphene relaxation time for different values of the chemical potential.

Standard THz-time-domain spectroscopy (TDS) measurements are performed to measure the graphene material parameters of the devices. The common thin film approximation for thin conducting layers connects the graphene material parameters with the obtained complex transmission [7]. Finally $\mu_c$ and $\tau_r$ are obtained by fitting the experimental values to the Kubo formula [8].

Simulations of the graphene antenna emission are performed by using CST [9] and then increasing the relaxation time progressively until an antenna resonance is visible. These simulations were performed for various values of the chemical potential.

The carrier mobility of graphene can be calculated for the evaluated values of $\mu_C$ and $\tau$ with [10]:
$\tau \approx \mu\, \mu_C/(ev_F^2)$. With the mobility $\mu$, the electron charge $e$ and the Fermi velocity $v_F \approx c_0/300$, with the speed of light $c_0$.

The simulation results as well as the experimentally obtained values are listed in table 1.

| Chemical Potential [eV] | Minimum Relaxation Time [ps] | Equivalent Carrier Mobility [cm$^2$V$^{-1}$s$^{-1}$] | Experiment/ Simulation | Figure of Merit ($\mu_C \cdot \tau^2$) [ps$^2$eV] |
|---|---|---|---|---|
| 0.2 | 0.7 | 35000 | Sim. | 0.098 |
| 0.4 | 0.5 | 12500 | Sim. | 0.1 |
| 0.6 | 0.4 | 6667 | Sim. | 0.096 |
| 0.8 | 0.35 | 4375 | Sim. | 0.098 |
| 1.0 | 0.3 | 3000 | Sim. | 0.09 |
| 0.26 | 0.046 | 1768 | Exp. | 0.00055 |
| 0.25 | 0.037 | 1480 | Exp. | 0.00034 |

Table 1: Minimum values of the graphene relaxation time for different chemical potentials, corresponding carrier mobility and figure of merit for simulated and experimental data

The minimum required relaxation time is always in the range of a few hundred fs and decreases with increasing $\mu_C$. This is due to the fact that more charge carriers are available for increasing chemical potential and thus the conductivity rises. For a clear antenna response graphene with both sufficiently high relaxation times and chemical potentials are required.

The experimentally obtained values of the relaxation time are significantly lower than the minimum required values indicated by the simulation results. Therefore the plasmonic resonance in such structures is expected to be strongly overdamped and a negligible antenna effect will be present.

We further investigated the dependency of the minimum relaxation time as a function of the chemical potential and could find a figure of merit ($\delta$) for the functionality of graphene based antenna devices:

$$\delta = \mu_C \cdot \tau_r^2 \qquad (2)$$

A plasmonic resonance is only present if $\delta \geq 0.1 \text{ ps}^2\text{eV}$. For the experimental data this would mean a necessary increase of a factor of 14 in relaxation time, by keeping $\mu_C$ constant comparing experimentally available materials and the necessary requirements of a reasonable performance graphene antenna.

Interestingly, the relaxation time has a greater influence on the antenna functionality than the chemical potential. Since the relaxation time is mostly influenced by the growth, transfer and the graphene environment, it is crucial to ensure a good graphene quality during the sample processing.

Further considerations clearly indicate that the plasmonic resonance in the experimental graphene samples is strongly overdamped, due to low relaxation times, i.e. fast scattering effects induced by sample imperfections (e.g. grain boundaries). A sufficient increase of $\tau_r$ or $\mu_C$ will result in a well pronounced antenna resonance.

With an antenna length of $L = 47$ μm and a width of $L = 5$ μm simulations show a well pronounced resonance for higher relaxation times at a frequency of $\nu_{res} = 0.4$ THz.

A gold antenna of same geometry was also measured and simulated for the sake of comparison. Both experiment and simulation show a well pronounced dipole resonance demonstrating the functionality of the model. The resonant frequency of this antenna occurs at $\nu_{res} = 1.2$ THz, which is a factor of 3 higher compared to the graphene antenna.

Thus a graphene antenna can be significantly smaller than their metallic counterparts by obtaining the same resonant frequency.

### III. Summary

In summary we present experimental and simulated minimum required material parameters for a graphene dipole antenna to emit THz radiation. It is shown that the plasmonic resonance strongly depends on the graphene properties. Furthermore a figure of merit is evaluated, which sums up the threshold parameters of $\mu_C$ and $\tau$. A reference gold antenna shows a resonance, which is a factor of 3 higher compared to a graphene antenna. This enables a significant downscaling of antennas by obtaining the same resonant frequency.

Acknowledgements: This work has been partially funded by the Spanish Ministry of 'Economía y Competitividad' under grant PCIN-2015-012, the German Research Foundation (DFG) under grants HA 3022/9-1 and LE 2440/3-1, the EU's Horizon 2020 research and innovation programme under grant agreement No. 736876, and the European Research Council (ERC, InteGraDe, 307311).

### References

[1]. S. Abadal, "Broadcast-Enabled Massive Multicore Architectures: A Wireless RF Approach," *IEEE MICRO,*, vol. 35, pp. 52-61, Oct., 2015.
[2]. P.R. West et al., "Searching for better plasmonic materials," *Laser and Photonics Review*, vol. 4(6), pp. 795-808, 2010.
[3]. Q.H. Park et al., "Optical antennas and plasmonics," *Contemporary Physics*, vol. 50(2), pp. 407-423, 2009.
[4]. S.Koenig et al., "Wireless sub-THz communication system with high data rate," *Nature Photonics*, vol. 7, pp. 977-981, 2013.
[5]. I. Llatser et al., "Comparison of the Resonant Frequency in Graphene and Metallic Nano-antennas," *AIP Conference Proceedings*, vol. 1475, 143, 2012.
[6] A. Cabellos-Aparicio, I. Llatser, E. Alarcón, A.Hsu & T. Palacios, "Use of Terahertz Photoconductive Sources to Characterize Tunable Graphene RF Plasmonic Antennas," *IEEE T. Nanotechnol.*, vol. 14, 390-396, 2015.
[7]. M.C. Nuss et al., "Terahertz time-domain measurement of the conductivity and superconducting band gap in niobium," *J. Appl. Phys.*, vol.70, pp. 2238-2241, 1991.
[8]. P.A. George et al., "Ultrafast Optical-Pump Terahertz-Probe Spectroscopy of the Carrier Relaxation and Recombination Dynamics in Epitaxial Graphene," *Materials Science and Engineering.*, vol.74, pp. 351-376, 2013.
[9] CST Microwave Studio [Online]. Available: http://www.cst.com
[10]. X. Luo et al., "Plasmons in graphene: recent progress and applications," *Materials Science and Engineering.*, vol.74, pp. 351-376, 2013.